# Electric Vehicle Enquiry (EVE) Pilot: 3-year operational data from a single electric car

Seun OSONUGA[a,1], Benoit DELINCHANT [a], and Frederic WURTZ[a]
[a] *Univ. Grenoble Alpes, CNRS, Grenoble INP\*, G2Elab, 38000 Grenoble, France*
ORCiD ID: Seun Osonuga https://orcid.org/0000-0003-4560-998X , Benoit Delinchant https://orcid.org/0000-0003-4296-0993 , Frederic Wurtz https://orcid.org/0000-0003-3117-9560

**Abstract.** This data paper presents the dataset from a study on the use of electric vehicles (EVs). This dataset covers the first dataset collected in this study: the usage data from a Renault Zoe over 3 years. The process of collection of the dataset, its treatment, and descriptions of all the included variables are detailed. The collection of this dataset represents an iteration of participative research in the personal mobility domain as the dataset was collected with low-cost commercially available equipment and open-source software. Some of the challenges of providing the dataset are also discussed: the most pertinent being the intermittent nature of data collection as an android phone and OBDII adapter were used to collect the dataset.

## 1. Introduction

In an era marked by accelerating environmental concerns and an urgent need for sustainable energy solutions, electric vehicles (EVs) are emerging as pivotal players in the ongoing global energy transition. As we collectively strive to reduce carbon emissions and dependence on fossil fuels, deploying EVs has gained prominence due to their potential to revolutionise transportation systems. In light of this, sharing operational data from personal electric cars stands as an essential stride towards understanding their utilisation dynamics and their potential for flexibility in our energy systems.

In many parts of the world, personal cars account for the most common forms of mobility. However, the most common forms of data on the use of personal vehicles are from travel surveys which are collected over typically a week and then extrapolated. In recent times, GPS technologies have been deployed to collect individual vehicle data over a long period (1). However, this process usually proves expensive (2). The need for cost-effective ways to collect personal vehicle data and the key role that electric vehicles will look to play in the future of not just mobility but also our energy systems give rise to the pertinence of this work.

The sharing of high-quality data is fundamental to research in both the mobility and energy domains. Many high-quality mobility datasets such as (3), although mentioned in the public domain are not exactly open. This is in contrast with REDD (4) and REFIT (5), popular datasets used in the energy domain. Therefore using a similar to the approach employed by (6), the EVE study aimed at gathering OBD-II sensor data in a participative manner. The distinctiveness of the dataset presented in this paper lies in its focus on personal electric car operations, encapsulating the intricacies of individual mobility patterns and vehicle behaviour within the broader energy landscape.

The inherent value of this dataset resonates on several levels. First, it uses a participative science approach, engaging with users as stakeholders in the research of EVs. This approach has the potential to enhance data availability and cultivate a sense of shared responsibility in driving the transition to cleaner transportation. Moreover, the creation of this dataset is characterised by a commitment to user privacy while enabling insightful analyses. As the study that produced this dataset potentially scales, it promises to transform into a wellspring of information, nurturing the growth of novel applications and strategies for mobility research.

---

[1] Corresponding Author: Seun OSONUGA, seun.osonuga@g2elab.grenoble-inp.fr .

In summary, this paper introduces a preliminary dataset from the Electric Vehicle Enquiry (EVE) study which seeks to explore the operational intricacies of personal electric cars in the European landscape. Specifically, this paper focuses on the pilot dataset from one user in the study and looks to direct potential re-users of the dataset to this one. The idea is to use the feedback from re-users to better pre-process the subsequent datasets tied to the EVE study.

## 2. Data and Methods

### 2.1. Data subject description

The electric vehicle in this study is a Renault Zoe 2014 Q90 (7), a mini-hatchback used primarily for personal activities by the user. It has an onboard type 2 AC 44kW charger and a 22-kWh battery. Its maximum speed is 135 km/h and can do 0 – 100 km/h in 13.2 seconds. The vehicle has no DC charger and its battery is air-cooled. More details on the specifications of the EV can be found in Table 1 below.

**Table 1.** Renault Zoe specifications table

| Specification | Value |
|---|---|
| Engine power | 65 kW / 88 hp |
| Maximum torque | 220 Nm |
| On-board charger AC | 43 kW (Type 2) |
| Battery capacity | 22 kWh |
| Battery pack nominal voltage | 360 Volts |
| # of battery modules | 12 |
| Battery module nominal voltage | 30 V |
| # of battery cells | 192 |
| Battery cell nominal voltage | 3.75 V |
| Battery module configuration | 8S2P |
| Maximum speed | 135 km/h |
| 0 – 100 km/h | 13.2 seconds |
| Curb weight | 1,405 kg |

The dataset is sourced from an individual EV user in the Auvergne-Rhone Alpes region of France over 34 months from October 2020 to August 2023. The user is a middle-aged man who works in the sustainability space and indicated that the energy transition as one of his major reasons for purchasing the vehicle. The electric vehicle was used mostly on the weekends (as seen on the driving heatmap in Figure 9) and on occasions during the week. The user also occasionally used it for longer trips to different areas of France as can be seen from the trip distance boxplot in Figure 3. The car was almost exclusively driven by a single user, however, there were occasional drives by one other individual. The electric vehicle was mostly charged at the user's house, but occasionally with chargers at his place of work and in public spaces. The user has a dedicated 3-kW plug in his garage that he installed when he bought the vehicle as he is the owner of his property.

### 2.2. Data typology

Table 2 below shows the description of the 139 variables in the dataset. It includes automotive data such as speed and distance covered, battery state data such as voltage and cell temperatures, charging data such as charging current and voltages, as well as other datasets such as external and internal temperatures, air-conditioning mode, and use of eco-driving mode. The data shared has not been resampled as it was the aim of the authors to share the data in its nearest to original form. As such there is a wide disparity in median timesteps of the variables ranging from 0.30 milliseconds to 16.73 seconds. As a result of this, the number of entries of the variables also varies greatly with variables with smaller timesteps having more values than the others. This disparity in timesteps is mostly likely by design as the sampling rate for the different variables is not the same due to their use by the car's computers. For example, cell temperatures were more closely monitored than external temperatures as cell temperatures had more to do with the safety of the car. Visualizations of select datasets are available in Section Data visualization below.

*2.3. Dataset construction*

*2.3.1. Participant mobilisation*

The study was carried out under the auspices of the Observatoire de la Transition Energetique (OTE-UGA) (8), an open science project at the University of Grenoble-Alpes funded by the French government. OTE-UGA looks to study various levers of the energy transition by carrying out multidisciplinary open and participative research. To achieve this, OTE-UGA has a panel of over 2500 volunteers who are interested in participating in open research linked to the energy transition. The data is the first dataset from a study on the use of electric vehicles. As the data from a single EV could be considered as information protected by the GDPR law, it was necessary to obtain the written consent of the user. The data flow and processing used by OTE-UGA from the selection of participants to the eventually sharing of the relevant data in open access can be found in Figure 1.

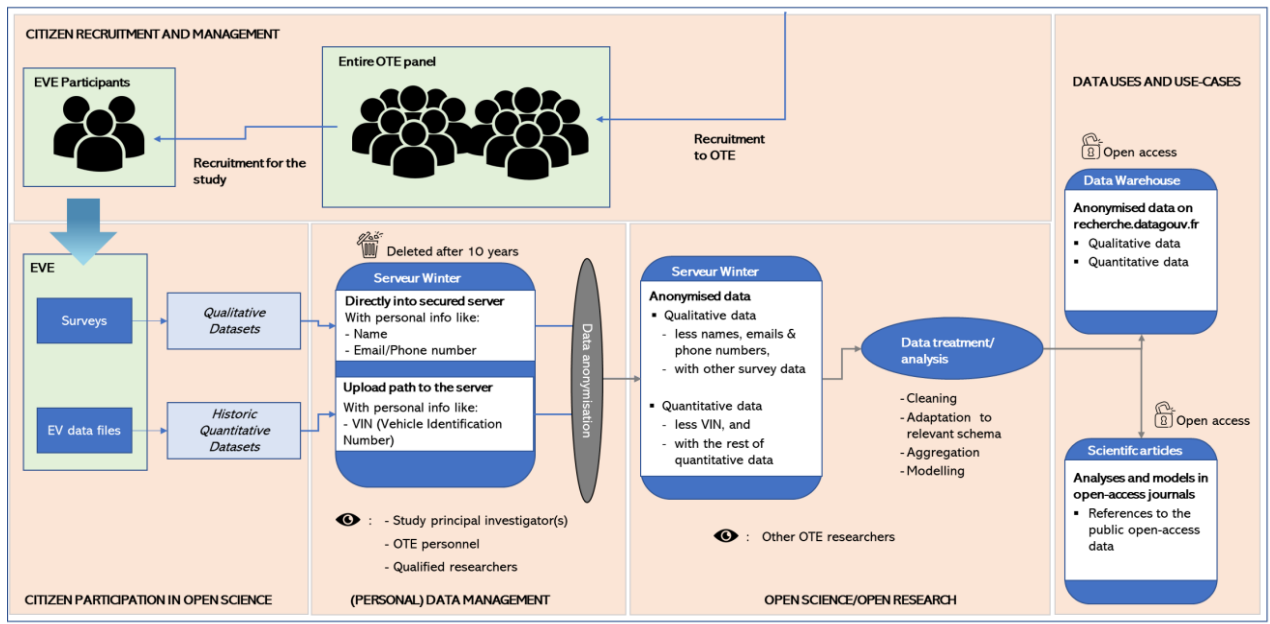

**Figure 1.** Process schema for the EVE Study.

*2.3.2. Data collection*

The quantitative data for this study was collected using openly accessible equipment and software. The data from the readings were collected from using a Bluetooth OBDII dongle (Konnwei KW902 ELM327 Bluetooth 3.0 OBDII Scanner) (9), open-sourced android application (CanZE) (10), and an android smart phone as shown in Figure 2.

The Bluetooth dongle was always connected to the OBDII port on the front console of the car. The collection of data and logging was trigged by the CanZE app whenever it was in Bluetooth range to the dongle and the car was in at least an accessories-start position. As such, the data collection and logging only occurred when the phone was in the car and mostly while charging in the home of the user. The data collected was amassed from the log files of the CanZE application. As the app was not originally conceived for the collection of data, a Python code was used to merge the logfiles and eventually create the datasets. Also, not all data variables are logged automatically by the CanZE app as a few of them such as module temperatures were only collected when the user views the relevant screen on the app.

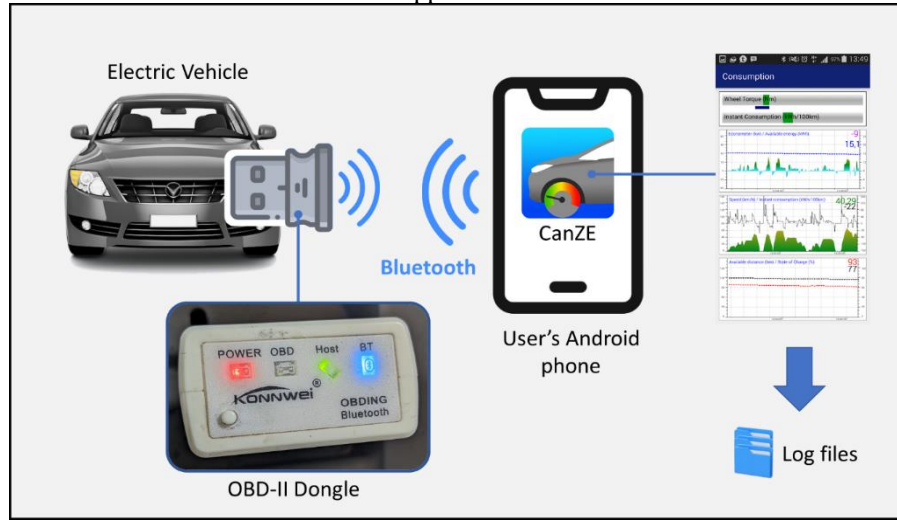

**Figure 2.** Quantitative data collection process.

The qualitative data about the use of the vehicle and the characteristics of the user were collected from the user through an online questionnaire available at the OTE study site (11). The sections of the questionnaire included: personal information, vehicle information, motivation towards the energy transition, use of the vehicle, and vehicle charging patterns. A list of the questions responded to by the user is available in Appendix A.

*2.3.3. Quality control methods*

Following the collection and merging of the data in log format, two broad forms of quality control were carried out: selection of the variables to be published, and the cleaning of the variables chosen. The first step was essential to ensure that no personal data captured in the logfiles was shared and that the most relevant variables were shared. The second step was carried out to remove the most obvious outliers and inaccuracies before sharing.

Initially, there were 5,015 unique variables in the logfiles, some of which were variable bugs from registration with the CanZE app. Table 2 shows frequency and type of all the 5,015 variables. The variables with more than 100 occurrences were selected and then divided into two groups: time-scale data and indicator data. From the indicator batch, only variables with more than 1 unique value were selected. From the time-scale data, we then chose datasets that matched with valuable data in five categories: drive-train, battery, charging, climatization & temperature and other datasets that seemed interesting.

**Table 2.** Frequency of all 5,015 variables

| Frequency of entries | Number of variables | Type of variable | |
| --- | --- | --- | --- |
| | | Time-series data | Indicator data |
| 1 - 10 | 4,721 | - | - |
| 11 - 100 | 32 | - | - |
| 101 - 1,000 | 16 | 10 | 6 |
| 1,001 - 5,000 | 75 | 48 | 27 |
| 5,001 - 50,000 | 143 | 118 | 25 |
| 50,000+ | 28 | 27 | 1 |
| Total | **5,015** | | |

*2.3.4. Data arrangement*

The data set is arranged in a file and folder format on recherche.data.gouv (12). The individual data timeseries are arranged in five files: Automotive, Battery, Charging,

Climatization and temperature, and Others, just as is laid out in Table 3. The files are in CSV format in the folders. As earlier mentioned the datasets has not been resampled, so there are uneven timesteps in most of them.

## 3. Results and Discussions

In this section, snapshots of certain data pieces, and some potential uses of the dataset are laid out. In addition, the section is ended with some of the challenges faced while collecting and analysing the data as well as some of the limitations of the dataset itself.

*3.1. Data visualization*

*3.1.1. Drive-train data*

The drive train folder contains speed, distance travelled, as well as braking and motor torques. As can be seen in Figure 3, the speed ranges from 0 till 130 km/h with the median speed at 21 km/h. And there is a median trip distance of 13 km. There are also occasionally longer trips of 200 km. This shows the typical use that is not daily for work transit but mostly for weekend errands and trips. As can be seen in Figure 4, the recorded distance covered by month is highly variable with less than 250 km covered in one month to more than 2000 km covered in another month.

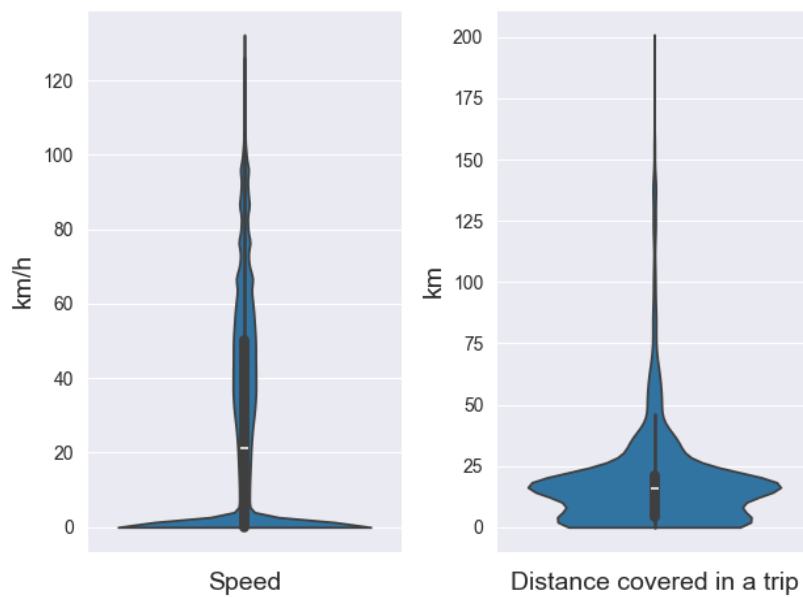

**Figure 3.** Violin plots of speed and distance covered in a trip

**Table 3.** Description of the 139 selected variables

| Data class | Variable | Units | Description | Median time-step |
|---|---|---|---|---|
| Automotive data | Speed | km/h | Vehicle speed as shown on the on-board speedometer | 1 |
| | Electric Motor Torque | N.m | Torque drawn from the electric motor | 1.26 |
| | Braking torque | N.m | Torque applied to the brakes | 1.26 |
| | Total vehicle distance | Km | Cumulative distance covered by the car as shown on the odometer | 22 |
| Battery Data | HV Battery Temperature | °C | The temperature of the entire battery pack | 0.30 |
| | HV LBC Voltage | Volts | The combined DC voltage of the entire battery pack (all 192 cells) | 1.28 |
| | HV LBC Current | Amps | The current drawn or sent to the entire battery pack from the electric motor. Electric battery to motor: Positive | 1.28 |
| | State of Charge | % | The state of charge of the battery pack as shown on the dashboard based on the usable battery capacity of 22 kWh | 11.95 |
| | Real State of Charge | % | The state of charge of the battery pack used internally by the electrical vehicle based on the total battery capacity of 25.92 kWh | 10.34 |
| | Available discharge energy | kWh | The amount of available energy in the battery based on energy obtained from charging | 10.54 |
| | Maximum battery input power | kW | The amount of power that the battery can allow in based on its State-of-Charge (SoC). This decreases as the SoC increases. | 10.50 |
| | Battery Health | % | Battery health calculated by an internal system algorithm | 10.40 |
| | Battery voltage 14v | Volts | Voltage of the 14v battery used to power car electronics such as windows, radio, etc. | 1.12 |
| | Module 1 – 12 Temperature | °C | The temperatures of all the 12 battery modules. | 10.59 |
| | Cell 1 – 96 voltages | Volts | The voltages of the 192 battery cells connected with pair parallel configuration (8s2p) | 1.4 |
| Charging data | Charging Plug detected | NA* | Indicator to denote whether a charger is plugged or not | 10.26 |
| | Mains current type | NA* | Detection of current type: Single-phase AC, Three-phase AC, or DC | 5.25 |
| | Mains phase 1, 2, 3 currents | Amps | Array of 3 RMS values of phase currents at the charger | 5.28 |
| | Mains AC line 1-2, 2-3, 3-1 voltages | Volts | Array of 3 RMS values of line voltages at the charger | 5.28 |
| | Mains active power consumed | kW | Real power consumed by the EV battery charger | 5.28 |
| | Available JB2 power for a battery charge | kW | Indicated the maximum power that can be supplied by the charging port that the car is connected to | 5.28 |
| | Mains ground resistance | Ohms | The resistance of the ground connection at the charger | 5.28 |
| Climatization & temperature data | IH_ExternalTemperature | °C | External temperature of the car | 2.48 |
| | In-car Temperature | °C | Temperature inside the car | 16.73 |
| | Relative Humidity | % | Relative humidity outside the car | 16.73 |
| | ILB_HPM_ClimLoopMode | NA* | Indicator that indicated the mode of the car's air conditioning system. Options include air-conditioning (1:AC mode), heat-pump (4:Heat Pump mode), and off (7:Idle mode) | 0.85 |
| Other | Start diag | NA* | Indicator that denotes the start of the communications between the CanZE app and the car's OBDII port. With some cleaning, it can be used to separate timeseries into data collection horizons | 300 |
| | IH_EcoModeRequest | NA* | Indicator denoting whether Eco mode on the climatization in the car was engaged or not | 5 |
| | Horizon | NA* | Derived variable from the logfiles that measures each horizon of constant data collection. Might include multiple trips if the phone was left in the vehicle between trips | NA* |

*NA\*:* Not applicable. Used to denote indicator data which has no unit of measure.

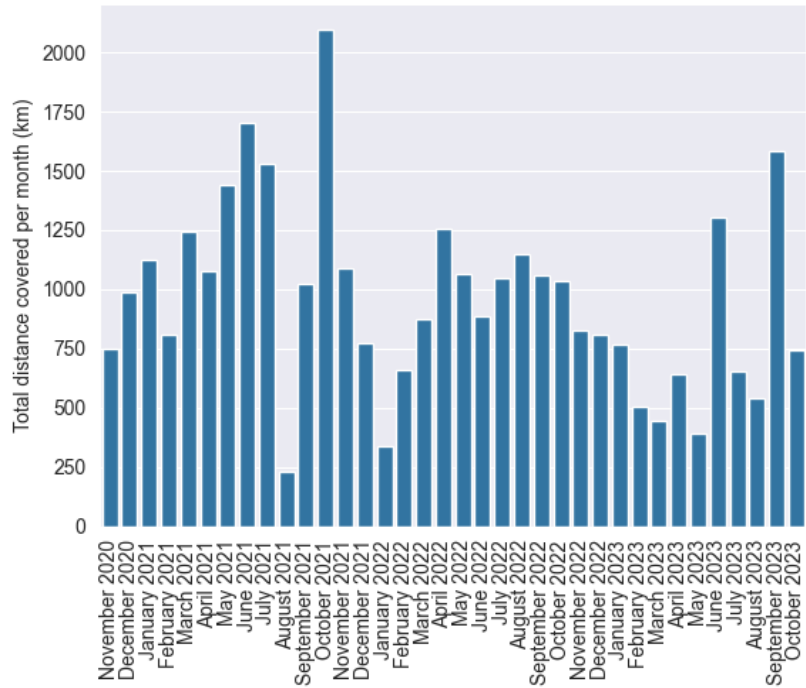

**Figure 4.** Total distance covered per month

### 3.1.2. Battery data

The battery folder contains several variables at pack level, temperatures at module level, and voltages at cell-pair level. Figure 5 shows violin plots of some variables at battery pack level: State of Charge (SoC), pack temperature, current and voltage with the most obvious outliers removed.

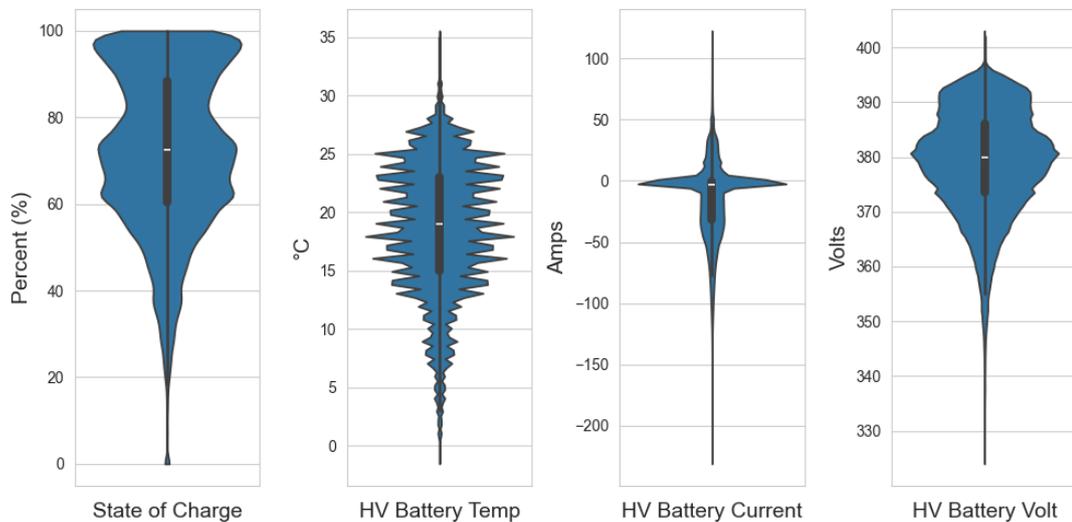

**Figure 5.** Violin plots of State of Charge, Battery Pack Temperature, Current, and Voltage

### 3.1.3. Charging data

The Charging data folder includes the phase currents, line voltages, power of EV charging port that the car was connected to amongst others. Figure 7 shows the distribution of different EV chargers used by the driver. As can be seen, most of the charging is done with 2 kW charger (mostly likely the charger in the user's home).

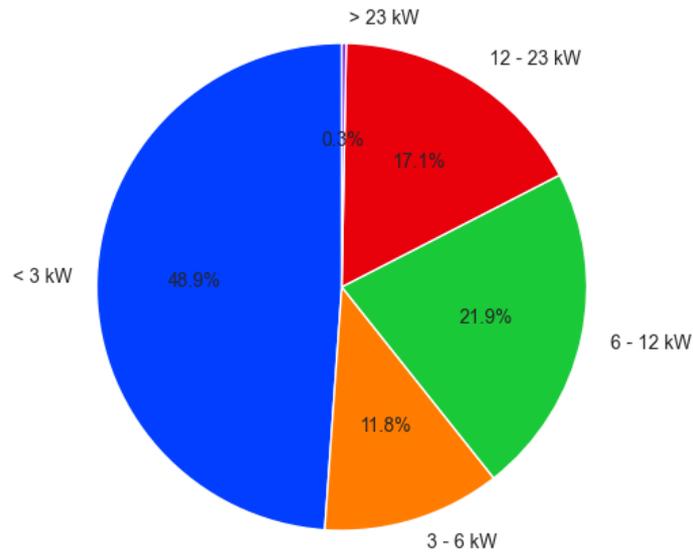

**Figure 7.** Pie-chart of the power of EV charging ports used by the driver.

### 3.1.4. Data availability

As highlighted in the Data collection sub-section, data was only collected when the user had his phone in the car. As such, there are periods for which there were no entries and seeming discontinuities in variables such as SoC and battery energy. Figure 8 shows the heatmap of data collection over the data collection period. As can be seen, there is almost no data between 00h and 07h, although the user indicated to frequently charge the car overnight.

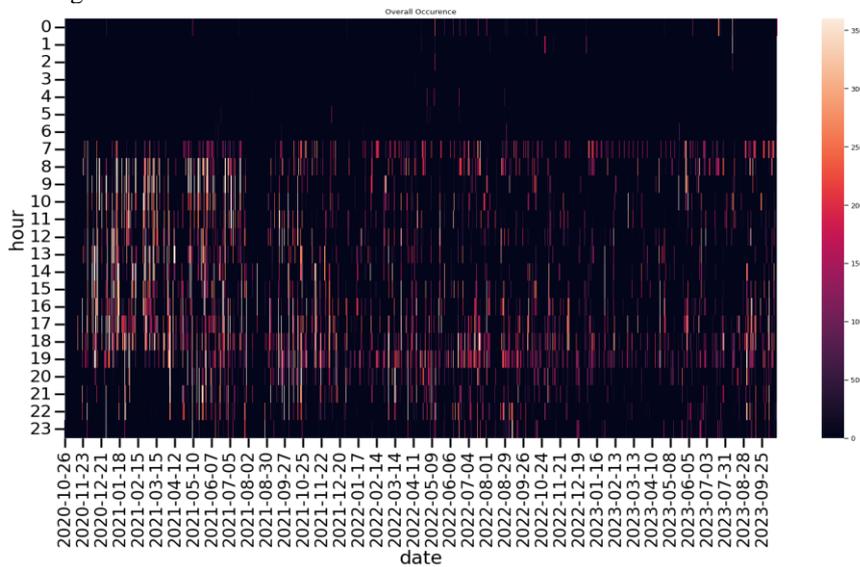

**Figure 8.** Heatmap of data availability per hour over the entire dataset

### 3.1.5. Vehicle use

As can be seen from Figure 9, the car is mostly driven during weekends, Monday mornings, and Friday evenings. This is a use that is typical of a personal vehicle used mostly for leisure activities and not for transit to work.

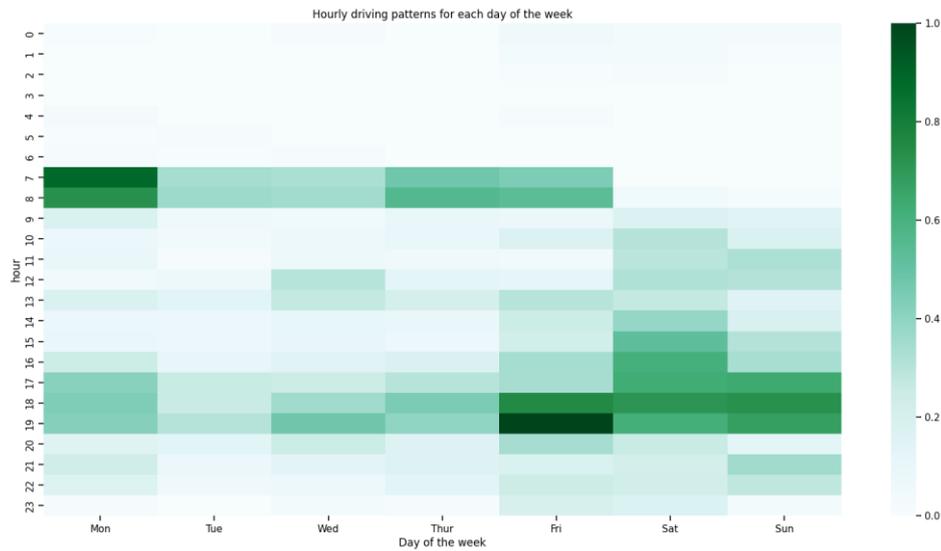

**Figure 9.** Normalised weekly hourly driving patterns over the 3 years

*3.2. Challenges and limitations*

In the preparation of the datasets there were a few challenges. The first was a lack of ready documentation on the logfiles obtained from the CanZe application. This is not part due to the fact that the primary reason for the app was data visualization on the mobile phone and not data collection to another source.

The data itself is also not without its limitations. Most pertinent among these is the lack of continuity in the flow of data. Because the collection was done on a mobile phone that connected to the OBDII dingle by Bluetooth, it was not possible to collect data when the user was not in the car. As a result, there are some holes in the data, for example between the start of a charging cycle and the end as these would map as time when the user gets to his home and then the next time he leaves.

There is also the limitation of representability. In France, 75% of car owners use them for transportation to work (13). However, this is not the case for the user who instead uses the car for mostly leisurely activities and personally errands. As opposed to this user's behaviour being reflective of a current typical user, it can be conceived as the proposed use for the future typical user as we prioritise more public transport and car-less city centres.

Another limitation to a much less extent is the lack of geographic data. The dataset includes only usage data and no location data which makes some forms of additional analysis more difficult and in some cases impossible. For example, characterisation of the driving style is subject to the type of roads and the behaviour of other road users, with geographic data, it is almost impossible to tell if the driver is moving very slowly on a suburban road or very quickly on a road in the city centre.

**4. Conclusions**

This paper lays out a dataset from a single EV user in France over two years. The dataset includes automotive data pieces like speed and distance, battery data pieces like SoC and voltage, and charging data pieces like charging power and mains powers. This paper has described the general data structure, data collection process, legal considerations for the collection, and some visualizations of certain datasets.

Prospects to this work include the extension of the pool of participants through the EVE study run by the OTE and potentially supplementing future versions with GIS data from Google Maps or Waze. Other formats for synthesising a larger pool of data have to be investigated as opposed to publishing datasets for each individual separately. Regarding the inclusion of GIS data into the EV data, ways of encoding that mask the actual location such as the pre-characterisation of roads and visited locations during data processing might be ways to further the work. There also exists a myriad of opportunities around open-hardware designs for the connection to the car with the possibility of some onboard storage to mitigate again data loss when the smartphone is removed from the vehicle.

**Acknowledgements**

This work was partially supported by the ANR project ANR-15-IDEX-02, the eco-SESA program (https://ecosesa.univ-grenoble-alpes.fr/), the Observatory of Transition for Energy (https://ote.univ-grenoble-alpes.fr/) , and a grant from the French government under the PIA, IRT Nanoelec, bearing the reference ANR-10-AIRT-05.

# Appendices

*Appendix A: List of questions to the user*

**Personal details (Adapted from OTE adhesion form)**
- Name
- Email
- Telephone number
- Year of birth
- Postal address and city
- Profession situation
- Distance between home and work
- Do you use multiple ways to get to work
- Comfort with IT tools

**Motivation for EV and the energy transition**
- Importance of energy transition when buying the car
- Reasons you bought the car
- Price range of the vehicle at purchase
- Time of purchase
- Maintenance history
- Major repairs

**Unique things for the car**
- Model, Brand, Year, and Trim of Car
- Primary use of car
  - Work
  - Travel
- Other uses of car
- Number of people who use the car as well and their ages
- Location of your EV data

**Actual EV OBD-II information**
- Battery and charging performance
  - SOC, Battery health, Voltage in time @ pack and cell levels
  - Charging times and locations
- Drive train settings
  - Speed
  - Braking
  - Distance covered
- Motor settings
- Auxilliary power use
  - Air conditioning
  - Heated seats
  - Entertainment system
- External conditions
  - Temperature, wind speed, precipitation
- Use information
  - Number of people in the car
  - Nature of trip (work, leisure, etc)

**Details of house for EV charging**
- Location of EV charging
- Type of house resident lives in
- Renter or Owner
- Electricity meter for the house
- Separate meter for your charger at home
- Type of EV charger
- Typical electricity bill
- Location of house (Urban, Suburbs, Rural)
- Estimate of EV users in resident's neighbourhood